\documentclass[a4paper,unpublished,noarxiv,onecolumn,11pt]{quantumarticle}
\pdfoutput=1
\usepackage[utf8]{inputenc}
\usepackage[english]{babel}
\usepackage[T1]{fontenc}
\usepackage{amsmath}
\usepackage{hyperref}
\usepackage{tikz}
\usepackage{amssymb,amsthm,mathrsfs,amsfonts}
\usepackage{physics}
\usepackage{enumerate}
\usepackage{color}
\usepackage{comment}
\usepackage{float}
\usepackage{caption}
\usepackage{subcaption}
\usepackage{adjustbox}
\usepackage[numbers,sort&compress]{natbib}
\usepackage{soul}

\newcommand{\be}{\begin{equation}}
\newcommand{\ee}{\end{equation}}
\newcommand{\bee}{\begin{equation*}}
\newcommand{\eee}{\end{equation*}}
\def\bal#1\eal{\begin{align}#1\end{align}}
\def\baal#1\eaal{\begin{align*}#1\end{align*}}
\def\bsub#1\esub{\begin{subequations}#1\end{subequations}}
\usepackage{graphics}
\usepackage{graphicx}

\definecolor{ForestGreen}{RGB}{34,139,34}

\begin{document}

\title{Contrary Inferences for Classical Histories within the Consistent Histories Formulation of Quantum Theory}

\author{Adamantia Zampeli}
\affiliation{Centro de Ciencias Matemáticas, Universidad Nacional Autónoma de México, 58190, Morelia, Michoacán, Mexico}
\affiliation{International Center for Theory of Quantum Technologies, University of
Gdánsk, Wita Stwosza 63, 80-308 Gdánsk, Poland}
\email{adamantia.zampeli@ug.edu.pl}
\orcid{0000-0002-1517-6127}

\author{Georgios E. Pavlou}
\affiliation{Institute of Applied and Computational Mathematics (IACM), Foundation for Research and Technology Hellas (FORTH), IACM/FORTH, GR-71110 Heraklion, Greece}
\email{gepavlou@iacm.forth.gr}
\orcid{0000-0003-4478-935X}

\author{Petros Wallden}
\affiliation{University of Edinburgh, School of Informatics, Informatics Forum, 10 Crichton Street, Edinburgh EH8 9AB, United Kingdom}
\email{petros.wallden@ed.ac.uk}
\orcid{0000-0002-0255-6542}

\begin{abstract}
In the histories formulation of quantum theory, sets of coarse-grained histories that are consistent obey the classical probability rules. It has been argued that these sets can describe the quasi-classical behaviour of \emph{closed} quantum systems, e.g. \cite{omnes1992consistent,Hartle1993}. Most physical scenarios admit multiple different consistent sets and one can view each of these as a separate context. Using propositions from different consistent sets to make inferences leads to paradoxes such as contrary inferences, first noted by Kent \cite{kent1997consistent}.
In this contribution, we use the consistent histories to describe a quasi-classical and macroscopic system to show that paradoxes involving contextuality persist even in the quasi-classical limit. This is distinctively different from the contextuality of standard quantum theory, where the contextuality paradoxes do not persist in the quasi-classical limit. Specifically, we consider different consistent sets for the arrival time problem of a (quasi-classical) ball in an infinite square well. For this setting, we construct two different consistent sets. We find the probabilities that each consistent set assigns to the simple question of whether the ball ever crossed the middle of the interval. We show that one consistent set concludes with certainty that the ball crossed it while the other consistent set concludes with certainty that it did not. Our results point to the need for constraints on the histories sets, additional to the consistency condition, to recover the correct quasi-classical limit in this formalism and lead to the motto `all consistent sets are equal', but `some consistent sets are more equal than others'.
 
\end{abstract}
\maketitle

\section{Introduction}
The consistent histories formulation of quantum theory \cite{gell1993classical, griffiths1984consistent, Omnes:1988ek, Omnes:1988em, Omnes:1988ej, omnes1992consistent} brings forward a proposal for dealing with the quantum mechanics of a closed system by replacing the observer-dependent process of quantum measurement with a \emph{consistency condition}. It is this condition that determines when a set of (coarse-grained) histories admits a classical description. However, the formalism allows for multiple consistent sets to exist for the same physical situation and the dominant view is that all these sets are to be treated equally\footnote{\label{multiplesets}An alternative view is that one consistent set (or a family of consistent sets) are more \emph{natural} (in some suitably defined sense) and should be given a preferred status in interpreting the theory. The physical principle that selects such consistent set(s) is still elusive even though certain proposals have been given \cite{anastopoulos1998preferred,2014FoPh...44.1195W,Halliwell_2017}}. Adrian Kent in ref. \cite{kent1997consistent} noted that, if we were to combine propositions belonging to different consistent sets, we would reach to ``paradoxes'' in the emergent classical logic such as contrary inferences. Two propositions represented by the projectors $P,\ Q$ at some moment of time are called {\it contrary} if $PQ=QP=0$ and $P< 1-Q$,\footnote{We can write $P_A P_B = P_A \wedge P_B :=P_{A\wedge B}$ only when $[P_A,P_B]=0$. If in addition $P_A P_B=P_BP_A=0$ we can write for the conjunction $P_{A\vee B} :=P_A \vee P_B=P_A + P_B$, see e.g. \cite{isham2001lectures}.} i.e. if they do not intersect and jointly do not span the full space at that moment.
\footnote{Compare with the definition of the {\it contradictory} propositions, which are those who do not intersect but cover the full space jointly at that moment, i.e. $PQ=QP=0$ and $P=1-Q$.} This reinforced the view, shared by proponents of the approach, that each consistent set is a context and that one should reason within a single consistent set each time. This would avoid the aforementioned paradoxes \cite{Griffiths:1997dt}. Contextuality, that is, reaching to different conclusions depending on the context in which the question is asked as well as the subsequent ``paradoxes'' that appear when one mixes contexts is a well-known and fundamental property of the (microscopic) quantum systems \cite{10.2307/24902153, peres1990incompatible, mermin1990simple, greenberger1989going}. It is also true though that such paradoxes should \emph{not} persist in larger scales when dealing, for example, with quasi-classical systems that have already decohered in some quasi-classical basis (e.g. the overcomplete basis of localized wavepackets) when subjected to local measurements.

In this contribution, we study a simple physical system at the quasi-classical limit within the consistent histories approach to show that the apparent ``contextuality'' arising from the analysis of different consistent sets is more paradoxical than the one of the ``standard'' quantum theory \cite{kent1997consistent}. In the latter, contextuality arises at a single moment of time; in our case, the different notion of contextuality arises because we speak of positions at different histories which involve sequences of different times.

We illustrate this through a concrete example based on the time-of-arrival problem for a coherent quantum state with positive momentum, which can represent, for instance, a tennis ball moving between two impenetrable walls. The coherent state is well-localised and chosen in such a way that we are clearly at the macroscopic limit. We construct two different consistent histories sets for the time-of-arrival problem and we show that they assign different probabilities to the question of whether the particle crossed the middle of the interval. Since there is no physical obstacle in the middle, our intuition suggests that the ball crosses the middle, so we should reject the set that gives different prediction. However, within the consistent histories approach there is no a priori way to distinguish between these two consistent sets since the consistency condition is not adequate to distinguish between the multiple consistent sets for one physical situation. This has led to the search for proper selection criteria for the consistent sets (see refs. of footnote \ref{multiplesets}).

\paragraph*{Overview of the results.} 
Before going to the details, we give here an intuitive explanation of our work. We summarise, at the end of this section, the three mathematical observations that lead to our result. 

Perhaps the easiest way to see the consistent histories formulation of quantum theory is to consider the Feynman path integral. Then the histories form the configuration space for the system for specific boundary conditions. The histories space is partitioned into disjoint subsets of histories. If all these coarse-grained histories pairwise have no interference, then this partition is a consistent set and the theory can give predictions (i.e. classical probabilities can be assigned). In the simplest case as the one we study here we have partitions with only two cells, where the one cell is the question we ask (e.g. whether the wave packet remains in the negative axis) and the other is its negation (e.g. whether the wave packet was at some point in the positive axis). This is the set-up of the time-of-arrival problem (or the quantum Zeno effect when we have discrete variables) which mathematically is equivalent with normal evolution with extra ``infinite walls'' - barriers - on the boundary of the region in question) \cite{Wallden:2008aa}
\bee
H_{\bar{\Delta}} = H + V_{\bar{\Delta}}
\eee
where $V_{\bar{\Delta}}$ is the potential of the infinite square well, see eq. \eqref{hamiltonian_well_restricted}. Now, in order for this partition to be a consistent set, the two cells of the histories space need to have zero interference. As was shown in \cite{Wallden:2008aa}, this leads to the condition $\bra{\Psi(\tau)}\Psi_r(\tau)\rangle=1$ together with a boundary condition, where $\ket{\Psi(\tau)}$ is the system unitarily evolved until the final time $\tau$ and $\ket{\Psi_r(\tau)}$ is the evolution of the system with the restriction that it did not leave the cell we ask for,  see eq. \eqref{consistency_cond} for a derivation. It is very important to note that in the consistent histories formulation there is no actual physical measurement taking place -- we are dealing with closed quantum systems without any observer -- and the ``barriers'' that are mathematically represented by projection operators is just a simpler way to describe the corresponding path integral that will tell us if the consistency condition is satisfied (and thus we can assign classical probabilities) or not (and thus this question involves quantumness and we cannot assign classical probabilities to those histories). Now, it is known that there are more than one consistent sets for the same histories space which means that we can define classical probabilities in different consistent sets and potentially get different predictions (probabilities) in each case. This happens when we have (at least) two different partitions of the histories space and when some coarse-grained histories from each partition form a zero measure cover of the histories space. Then it was shown in \cite{2014FoPh...44.1195W} that these propositions are contrary. In the case they both have probability one to happen this constitutes a problem for the quantum theory which becomes severe when we deal with the quasi-classical limit as we discuss in this work.

Coming back to our example, we note that mathematically the consistency of the set depends on the final time $\tau$ and the restricted evolution imposed by the region we consider. In particular, the set is still consistent even if, in some earlier time, this condition was not satisfied. Therefore, to find a consistent set that contradicts our classical intuition, we simply need to find a quasi-classical system that after some time evolves to the same state as if it was restricted to a smaller region, perhaps by ``bouncing'' and returning to the same place. We achieve this by looking at a free wave packet in an infinite square well localised in the negative axis moving with positive average momentum. The question asked (restriction) is whether it crosses to the positive axis. The significance/effect of the infinite square well is that the ``unrestricted'' evolution of the wave packet bounces back and coincides after some time with the restricted evolution. We have now just constructed a consistent set that concludes that the particle never crosses the axis, even though it is a quasi-classical object moving towards there. One might be led to believe that  for this question the consistent histories approach cannot recover the ``correct'' answer. However, we note that it is easy to find another consistent set that actually gives the intuitive answer: we can consider a partition such that the restricted evolution of the wave packet is in a small region that follows the classical path, as is shown in figure \ref{fig:arrival_time2}.  The surprising conclusion is that we have constructed two consistent sets, disagreeing about a classical property (passing into the positive axis/position) of a classical object. We recall at this point that in both sets, the ``restrictions'' come from the question we ask, i.e. by simply partitioning the history space in two different ways. There is no physical difference in the two scenarios. The histories space (all possible paths) and the dynamics (that are encoded in the unrestricted Hamiltonian that defines the decoherence functional) are identical.

Our results are based in the following three mathematical observations:

\begin{enumerate}
\item As shown in Refs. \cite{sorkin_exercise_2007,Wallden:2008aa}, the conditional state of the system for the history corresponding to the question ``does the particle stay in a time-dependent region $\Delta (t)$ for all times?'' is obtained, through the quantum Zeno effect, by evolving the initial state with the original Hamiltonian supplemented by an infinite constraining potential $V_\Delta$ for $\Delta (t)$:
\be
C_{\Delta} = \exp (-i H_\Delta t), \quad H_\Delta = H + V_\Delta
\ee
where $C_\Delta$ is the class operator corresponding to the set of histories that stay in $\Delta$.

\item The consistency of the two-history set $(C_\Delta, C_{\bar{\Delta}})$ is equivalent to having unit overlap of the restricted and unrestricted evolutions:
\bal
&D_{\Delta,\bar{\Delta}} = \Tr [C_\Delta \rho C_{\bar{\Delta}}^\dag] = 
\Tr [\exp (-i H_\Delta t) \rho \exp (i H_{\bar{\Delta}} t)^\dag] = \nonumber \\
&\Tr[\exp (-i H_\Delta t) \rho \exp (i H t)] -1 = 
\braket{\psi_r}{\psi} -1
\eal
where $\rho$ is the initial density matrix and $\psi_r$ the state evolving with the restricted propagator.

\item The above histories for two distinct regions $\Delta_1 (t)$ and $\Delta_2 (t)$, could be contrary. Specifically, if there exists a time $t\in [0,\tau]$ such that $\Delta_1 (t)\cap\Delta_2(t)=\emptyset$, i.e. if the two regions are disjoint for some moment of time within the time interval considered, then the corresponding two consistent sets lead to contrary inferences\footnote{In other words, in that moment $t$, each of the two consistent sets asserts with certainty that the system is in the region $\Delta_i(t)$, but depending the set, it concludes with certainty that the system is at a disjoint region.}. What our work does, is to find two such regions $\Delta_1(t),\Delta_2(t)$, for a ``classical'' system that both satisfy the consistency condition of point 2 above and that are disjoint for one moment of time.
\end{enumerate}

\paragraph*{Related works.} The interpretation of consistent histories has been a heated topic (indicatively, examples of this dialogue can be found on the critical accounts of  \cite{okon_consistency_2014,okon_measurements_2014,bassi_decoherent_2000} that were followed by responses of proponents \cite{griffiths_consistent_2015,griffiths_consistent_2000}). Our work intends to add one more dimension in this scientific exchange strengthening the argument for supplementing the consistency condition with a ``set-selection'' criterion. 
This issue was first raised by Dowker and Kent \cite{dowker1996consistent} and later by Kent \cite{kent1997consistent} who noted that the problem of contrary inferences is the result of a contextuality particular only to the histories formalism. He later gave a set-selection criterion to restrict ``ordered consistent sets'' \cite{MR1644164, kent1997consistent, Kent:1996kb}.  
In \cite{2014FoPh...44.1195W} it was argued that contrary inferences originate from the existence of zero covers on the histories space and brought forward another set-selection criterion that restricts attention to preclusive consistent sets. In \cite{anastopoulos1998preferred}, Anastopoulos connected the predictability of a consistent set with the persistence in time; 
while recently, Halliwell proposed to classify multiple consistent sets on whether or not there exists any unifying probability for combinations of incompatible sets which replicates the consistent histories result when restricted to a single consistent set \cite{Halliwell_2017}. 
Gell-Mann and Hartle \cite{gell-mann_quasiclassical_2007} and Riedel, Zurek and Zwolak \cite{riedel_objective_2016} on the other hand gave set-selection criteria that were motivated by  some physical principle (e.g. thermodynamic considerations) or induced by the environment respectively. 

In the preclusive consistent sets criterion of \cite{2014FoPh...44.1195W}, histories that are precluded (have measure zero) have a special role, since such histories and their subsets are not allowed to happen in any ``legitimate'' (i.e. preclusive) consistent set. Precluded histories play crucial role in another histories formulation, called co-events formulation or anhomomorphic logic \cite{sorkin_quantum_1997,sorkin_exercise_2007}, that interprets the decoherence functional (or quantum measure \cite{sorkin_quantum_1994}) differently. Our results may have consequences in that approach too, but analysing this is left for a future work.

The example we introduce, that within the consistent histories illustrates our point (existence of contrary inferences for classical systems), is based on the well-known time-of-arrival problem for a free particle \cite{allcock_time_1969}. The time-of-arrival type of questions constitute prototypes for dealing with questions that involve a non-trivial temporal aspect (c.f. time extended propositions). Standard quantum theory cannot answer such questions easily, possibly due to the fact that time is not an observable (no self-adjoint time operator), while it only appears as an external parameter marking the evolution in Schr\"odinger equation. 
Within the (consistent) histories framework temporal questions without external observers (closed systems) can be addressed naturally. For example, time-of-arrival histories are special cases of spacetime coarse-grainings of the histories space \cite{Hartle:1991qg}.

There have been many works addressing temporally extended questions within the consistent histories formulation. For instance Halliwell and Zafiris in \cite{1998PhRvD..57.3351H} showed that coupling of the particle with a thermal bath allows the definition of coupling-dependent probabilities for the time-of-arrival problem; Anastopoulos and Savvidou in \cite{anastopoulos_time--arrival_2006}  studied the issues of the definition of a temporal probability density in the context of Copenhagen interpretation and of histories formalism and attempted to construct positive operator valued measurements by considering them smeared in time, while 
in \cite{Halliwell:2006aa,Wallden:2007aa,Wallden:2008aa} it was investigated the time-of-arrival problem for closed quantum systems in non-relativistic quantum theory and in time-reparametrisation invariant theories.

The time-of-arrival histories are also related to the quantum Zeno effect \cite{Misra_1977,facchi2001zeno} which states that the evolution of a system ``freezes'' under continuous observation (see for example \cite{Wallden:2007aa}). In infinite-dimensional projection operators, as the one we are dealing with in this work, it has been shown that the evolution is restricted to a subspace of the Hilbert space \cite{FACCHI200012}. These cases usually refer to spatial projections but they also accommodate more general questions regarding the assignment of some property of the quantum system. In this case, the projector confines the system in some range of the spectrum of the corresponding observable.

These ideas have also been used in the context of quantum gravity for the definition of observables. Quantum gravity is a genuinely time-reparametrisation invariant theory and treats time as a dynamical degree of freedom, in contrast to quantum theory in which time is just a background parameter. The time-of-arrival type of questions in quantum gravity can be used to define the observables of the theory, by asking whether the system is in some region of the configuration space (or equivalently in a subspace of the Hilbert space) \cite{grot_time_1996,rovelli_partial_2002}. Consistent histories formalism offers again a suitable framework to work with this type of questions in quantum gravity since not only can naturally be formulated, but also they can naturally describe closed systems \cite{isham_continuous_1998,Wallden:2008aa,Christodoulakis:2011zz}.


\paragraph*{Organisation of the paper.}
This paper is organised as follows: In section \ref{time_arrival} we review the time-of-arrival problem in the consistent histories approach and discuss the definition and appearance of contrary inferences. In section \ref{consistent_sets}, we define the suitable consistent sets; we give a detailed account of the assumptions that this construction is based on and demonstrate that the consistency condition is indeed satisfied for both coarse-grainings. After that, we discuss and conclude in section \ref{discussion}. Finally, in Appendix \ref{Preliminaries} we review the histories formalism while in Appendix \ref{quasi-classical} we give details about the quasi-classical approximation and the numerics.

\section{The time-of-arrival problem in the consistent histories approach and contrary inferences}
\label{time_arrival}

\begin{figure}[t!]
\centering
\begin{subfigure}[t]{0.45\textwidth}
\centering
\includegraphics[height=3.5cm,width=6.5cm]{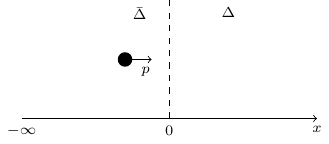}
\caption{Time-of-arrival problem for a free particle}
\label{fig:arrival_time}
\end{subfigure}
\begin{subfigure}[t]{0.45\textwidth}
\centering
\includegraphics[height=4cm,width=6.5cm]{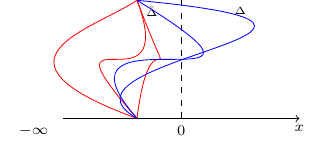}
\subcaption{Histories in the time-of-arrival problem}
\label{fig:arrival_time_histories}
\end{subfigure}
\caption{Left Panel: Instant of the coarse-graining for the time-of-arrival problem for a free particle. The question is whether the particle will have for at least some moment $t$ positive position coordinate. A natural coarse-graining is to select the intervals $\Delta (t)= [0, +\infty )$ and $\bar{\Delta} (t)= (-\infty,0]$ as shown here. \newline Right Panel: Various histories in the time-of-arrival problem in a specific partitioning -- red paths remain always in the region $\bar{\Delta}$, blue paths have being in $\Delta$ at some moment of time.}
\label{fig:arrival_time_pictures}
\end{figure}

The question ``what is the probability that the particle will be (``arrive'') in some region (range) $\Delta$ during some interval $[0,\tau]$?” is usually referred to as the time-of-arrival problem. The set-up of the problem, shown in Figure \ref{fig:arrival_time}, is that of a free particle with support in the interval $\Delta$ and its complement $\bar{\Delta}$ defined respectively by $\Delta = [0, +\infty )$ and $\bar{\Delta} = (-\infty,0]$, which evolves under the action of the free one-dimensional Hamiltonian. The particle is initially located in the negative axis with positive average momentum and it is thus expected to cross the surface $x=0$ at some moment $t \in [0,\tau]$. We consider histories that frequently check if the particle is still in $\bar{\Delta}$, depicted in red colour in Figure \ref{fig:arrival_time_histories}, by applying very frequently the continuous projection operator  
\be\label{proj_oper_delta}
\bar{P}= \int_{\bar{\Delta}} dx \ketbra{x}{x}.
\ee 
At the limit of infinitely many projections $N \rightarrow \infty$ and at infinitesimal time interval $\delta t=T/N \rightarrow 0$, it has been shown that the evolution of the system is restricted in the subregion $\bar{\Delta}$ with evolution propagator \cite{Wallden:2008aa}
\be\label{restricted_prop}
g_r (t) = \exp (-i H_r t) \bar{P}
\ee
where $H_r$ is the restricted Hamiltonian given by 
\be\label{restricted_ham}
H_r \equiv  \bar{P}H\bar{P}
\ee
and defined on the proper subspace $\mathcal{L}^2 (\bar{\Delta})$ with domain $D(H_r) = \{ \Psi \in \mathcal{L}^2(\mathbb{R}) | \partial_x \Psi \in \mathcal{L}^2(\mathbb{R}), \Psi (\partial \bar{\Delta} =0)  \}$. This can alternatively be written as \cite{FACCHI200012}
\be\label{hamiltonian_well_restricted}
H_r= - \frac{\hbar^2}{2M}\frac{d^2}{dx^2} + \bar{V}(x) , \quad 
\bar{V} (x) =  
\begin{cases}
0, & x \in \bar{\Delta},\\
\infty , & \textrm{otherwise}
\end{cases}
\ee
where $M$ denotes the mass of the particle and $\hbar$ the reduced Planck constant, which means that the particle behaves as if it were confined in $\bar{\Delta}$ by rigid barriers when one makes infinitely many measurements (projections with \eqref{proj_oper_delta}).\footnote{We will call the walls induced by infinitely many projections ``barriers" while we keep the term ``wall'' for the boundaries of the infinite square well in which all the dynamics happens.} This construction has also been called ``Zeno box'' \cite{SCHULMAN2002823}, because it is the analogue of the quantum Zeno effect for continuous variables systems. We emphasise that this is only a mathematical equivalence, since in our considerations there are no measurements, the projections arise in our effort to compute the amplitude of a particular (restricted) set of histories.


To express this problem in the consistent histories approach, we consider the coarse-grained alternative histories 
\bsub
\bal
&h =  \{ x (t) \  | \  x (t)  \in \bar{\Delta} \  \forall t \in [0,\tau]\}, \\
&\bar{h}  =\{ x (t) \  | \  x (t)  \notin \bar{\Delta} \text{ for some } t \in [0,\tau]\}.
\eal
\esub
These histories define the set $\mathcal{C} = \{h , \bar{h}\}$ and form a two-histories partition of the histories space $\Omega$ since they satisfy the relations
\be
h \cup \bar{h} = \Omega, \quad h\cap \bar{h} = \varnothing.
\ee

Note that while this pair of histories form a partition, in general these histories would \emph{not} decohere and this partition would \emph{not} constitute a consistent set. Indeed in order to assign probabilities to each of the histories, the set of the alternative histories $\mathcal{C}$ must be consistent, i.e. the non-diagonal terms of the decoherence functional must vanish. In this set-up this condition takes the form
\begin{eqnarray}\label{consistency_cond}
D(h, \bar{h})=0& =& \bra{\Psi_0}g^\dagger_r(t,t_0)\left(g(t,t_0)-g_r(t,t_0)\right)\ket{\Psi_0}\nonumber\\
&= &\bra{\Psi_0}g^\dagger_r(t,t_0)g(t,t_0)\ket{\Psi_0}-\bra{\Psi_0}g^\dagger_r(t,t_0)g_r(t,t_0)\ket{\Psi_0} \nonumber\\
&=& \bra{\Psi_0}g^\dagger_r(t,t_0)g(t,t_0)\ket{\Psi_0}-\bra{\Psi_0}\bar{P}\ket{\Psi_0}\nonumber\\
\bra{\Psi_0}\bar{P}\ket{\Psi_0} &=& \bra{\Psi_0}g^\dagger_r(t,t_0)g(t,t_0)\ket{\Psi_0}.
\end{eqnarray}
where we have used that i) $g_r(t,t_0)\ket{\Psi_0}$ is the evolution restricted in $\bar{\Delta}$ and corresponds to history $h$, ii) $\left(g(t,t_0)-g_r(t,t_0)\right)\ket{\Psi_0}$ is the evolution that is not always in the region $\bar{\Delta}$ and corresponds to the history $\bar{h}$ and iii) if $\ket{\Psi_0}$ is initialised in the region $\bar{\Delta}$ then we have $\bra{\Psi_0}\bar{P}\ket{\Psi_0} =1$. This relation, with the above assumptions therefore gives:
\be\label{consistentcond2} 
\bra{\Psi_0} g_r^\dag (t,t_0) g (t,t_0) \ket{\Psi_0} = \bra{\Psi_r(t)} \ket{\Psi(t)}=1
\ee
This condition states that, for the set of histories to be consistent, the overlap between the full and the restricted propagator at some time $t$ must be equal to unity (or more generally, if the state was not localised in $\bar{\Delta}$, equal to the overlap at the initial time $t_0$).\footnote{We note here that the calculation of the overlap between the full and restricted wave functions involves the inner product of two functions defined in different Hilbert spaces; $\mathcal{L}^2 (\Delta) \neq \mathcal{L}^2 (\Delta_{2L})$. Thus, one has to extend the definition of the restricted wave function on the full Hilbert space of the problem by embedding the restricted wave function in this larger Hilbert space. Therefore we have 
\be
\mathcal{L}^2 (\Delta_{2L}) = \mathcal{L}^2 (\Delta) \oplus \mathcal{L}^2 (\bar{\Delta})
\ee
where $\bar{\Delta}=\Delta_{2L} \backslash \Delta$ is the complement of $\Delta$.} 
In this case we have a consistent set and classical probabilities can be assigned. 
Thus the consistency condition breaks into this expression and the boundary condition 
\be\label{boundcond}
\braket{x}{\Psi_0}_{\partial \bar{\Delta}}=0
\ee
These are the conditions we will later use to construct our consistent sets and to show that they lead to contrary inferences. This is also possible because we can identify coarse-grained histories with propositions (and operators). 

A proposition that happens with probability one is assigned truth value $\textrm{True}$. It is not hard to see that if one has two contrary propositions (say $P,Q$) that both occur with probability one, it means that they both have truth value $v(P) = v(Q) = \textrm{True}$, something that leads to logical paradoxes. By the law of inference (modus ponens), if a proposition $P$ is assigned truth value $\textrm{True}$, so does any proposition that contains $P$. However, we also have that $1-Q>P$ so that the proposition $1-Q$ should also take value $\textrm{True}$ which leads to the situation that both $v(Q)$ and the negation $v(1-Q)$ are $\textrm{True}$. To reach this paradox we had to make an inference (thus the term ``contrary inferences''). Interestingly, in consistent histories, one can get two contrary propositions having probability one (each defined in a different consistent set).

It is worth pointing out, that this type of paradox, that leads to contrary inferences, is not present in standard ``single-time'' quantum theory. We can prove this by contradiction. Assume propositions $P,Q$ that are contrary, and thus $P<1-Q$. In standard quantum theory, these propositions are projection operators. Let us consider a state $\rho$ such that $\Tr (Q\rho )=1$, implying that we should assign True to the proposition $Q$. Then it follows that $\Tr ((1-Q)\rho)=0$, that also means that $\Tr (P\rho)=0$, since $P$ is a projection in a subspace of $(1-Q)$. This means that if $Q$ is assigned True with certainty, then $P$ must be assigned False with certainty. This simple argument fails in the histories setting because the consistency of the sets depends on the selection of the coarse-graining on the configuration space $\Omega$ in contrast to the case of the single-time propositions which are independent of this selection.

In \cite{2014FoPh...44.1195W} it was highlighted that one can view the origin of contrary inferences in the existence of a ``two-histories zero measure cover'' of histories $\Omega$, which is a pair of (coarse-grained) histories $h_1, h_2$ each of them having zero measure $\mu (h_1)=0, \ \mu(h_2)=0$ and covering $\Omega$ in the sense that $h_1 \cup h_2 = \Omega$. Let us denote $\bar{h}$ the negation of the history $h$.
The existence of two-histories of zero measure cover means that we can find two consistent sets of histories $C_1 =\{ h_1, \bar{h}_1 \}$ and $C_2 =\{ h_2, \bar{h}_2  \}$ such that the histories $h_i, \bar{h}_i$ satisfy the relations 
\begin{align}
&\mu (h_1) = 0, \ \mu(\bar{h}_1)=1, \\
&\mu (h_2) = 0,  \ \mu(\bar{h}_2)=1,\\
&\Omega=h_1\cup h_2
\end{align}
We can see that $\bar{h}_1 \cap \bar{h}_2 = \varnothing$ which means that the two histories $\bar{h}_i$ have no common fine-grained history and are contrary propositions. Since both $\bar{h}_i$  occur with probability one, they should be assigned truth value $\textrm{True}$ leading to contrary inferences. A simple physical example that arises is the three-slit experiment \cite{2014FoPh...44.1195W,sorkin_quantum_1997}. Note, however, that in the three-slit experiment, the histories (and the consistent sets) ``retrodict'' the slit that a photon (or quantum particle) passed from. Obtaining a contextual description for the history of a system in the quantum realm is something widely accepted. What we show in this paper is that in the consistent histories, such contextuality persists for quasi-classical systems.


\section{Contrary inferences in a quasi-classical system}\label{consistent_sets}

\subsection{Quasi-classical and macroscopic approximation}
\begin{figure}[ht!]
\centering
\begin{subfigure}[t]{0.45\textwidth}
\centering
\includegraphics[height=4cm,width=6.5cm]{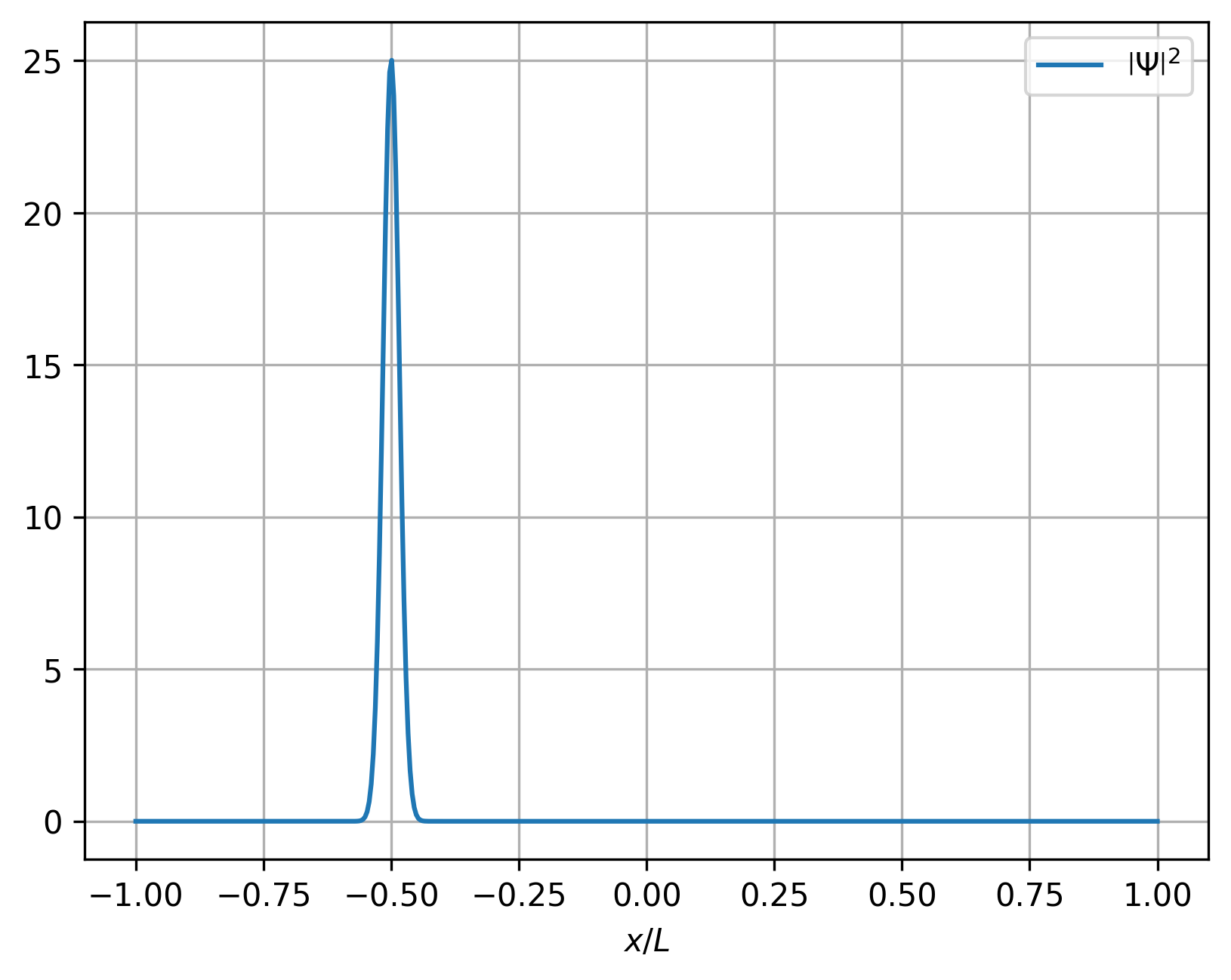}
\subcaption{Absolute square of wave functions $\Psi$ for $t=0$. The probability distribution of the initial coherent state has a Gaussian form in the position space.}
\label{fig:sigma10_psi}
\end{subfigure}
\begin{subfigure}[t]{0.45\textwidth}
\centering
\includegraphics[height=4cm,width=6.5cm]{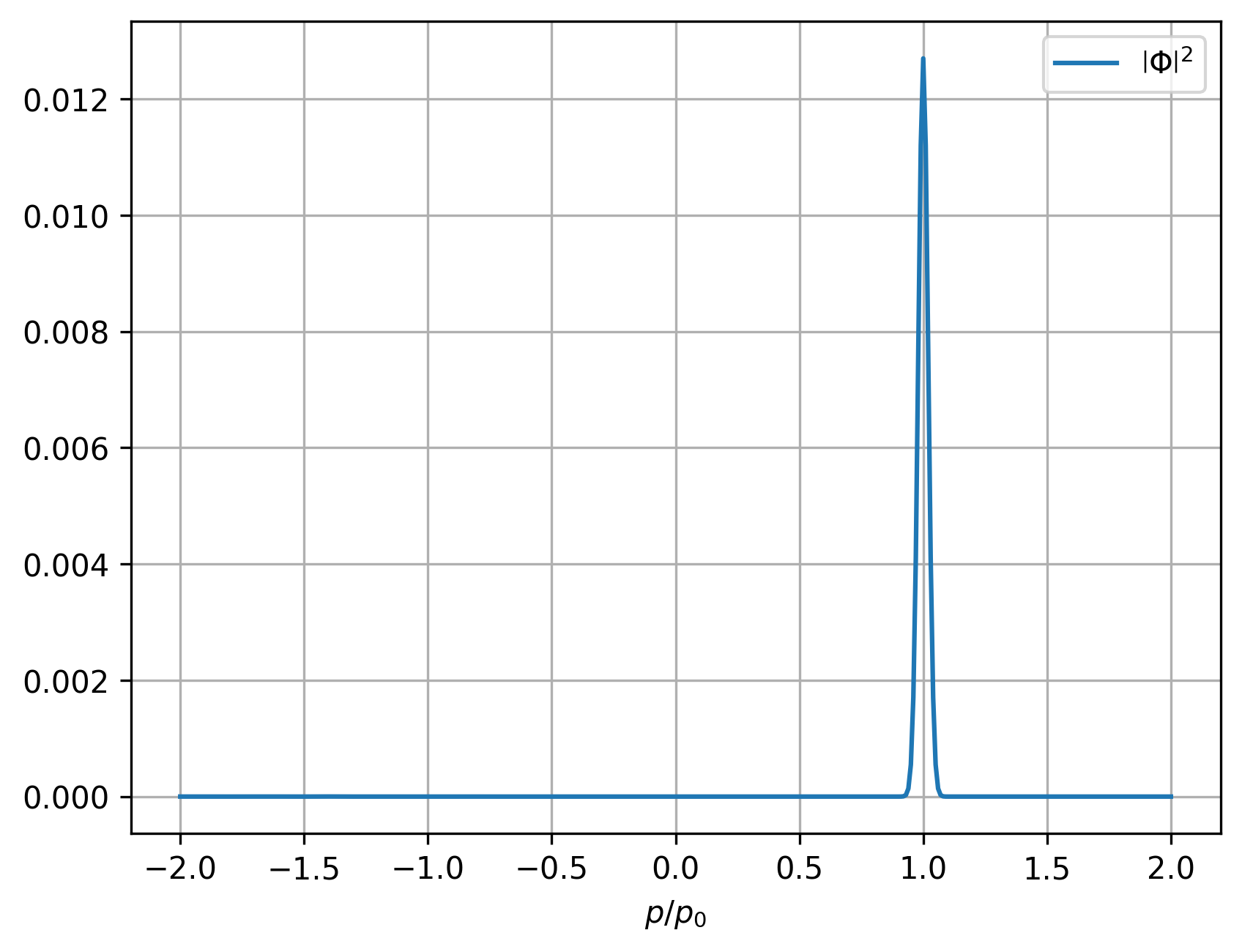}
\subcaption{Absolute square of the Fourier transform of the wave functions $\Psi$ for $t=0$ denoted as $\Phi$. The quantum state in the momentum state is a Gaussian wave packet centred in the initial momentum $p_0$}
\label{fig:momentum_space}
\end{subfigure}
\caption{Probability distribution of the coherent state in position and momentum space at $t=0$. The parameters were chosen as follows: $n_0 = 201$ and $\sigma_0 = 20$. The phase $\varphi_0$ is selected such that the wave function is localised at $x/L = -1/2$, as demonstrated in the figure.}

\label{initial_prob_distr}
\end{figure}
In this section we will show that there exist (at least) two different \emph{consistent} sets, each of which gives a different answer to the time-of-arrival question. We explicitly \emph{construct} them and show that they satisfy the consistency conditions \eqref{consistentcond2} and \eqref{boundcond}. 
We consider a coherent state to represent the particle in an infinite square well $[-L,L]$ with Hamiltonian
\be\label{hamiltonian_well}
H = - \frac{\hbar^2}{2M}\frac{d^2}{dx^2} + V(x), \quad 
V (x) =  
\begin{cases}
0, & \abs{x}< L,\\
\infty , & \abs{x} \geq L
\end{cases}
\ee
This Hamiltonian is a self-adjoint operator on the vector space $\mathcal{L}^2 (\Delta_{2L})$ and describes the free evolution in $\Delta_{2L} = [-L,L]$.
To construct the coherent state, we start with a wave function of the form
\be \label{expansion}
\Psi(x,t)= \sum_{n=0}^\infty c_n u_n (x) e^{-i E_n t/\hbar}
\ee
which is a linear superposition of the energy eigenstates
\be\label{stationary_states}
u_n = \sqrt{\frac{1}{L}} \sin \frac{(n+1) \pi (x+L)}{2L}
\ee 
with energy eigenvalues
\be
E_n = \frac{\hbar^2 \pi^2 (n+1)^2}{2 M (2L)^2} = (n+1)^2 E_0,
\ee
where $E_0$ is the energy of the ground state, while the momentum eigenvalue is given by
\be
p_n = (n + 1)\frac{{\pi \hbar }}{{2L}}.
\ee
In \cite{2000PhRvA..61c2107F,fiset2014equivalent} coherent states for the infinite well were constructed, known as Gaussian Klauder coherent states. These are given by the following selection for the coefficients in equation \eqref{expansion}:
\be
\label{c_n0}
c_n (n_0,\sigma_0,\phi_0) = \frac{e^{-\frac{(n-n_0)^2}{4\sigma_0^2} -in\phi_0}}{\sqrt{N(n_0,\sigma_0)}}
\ee
where 
\be
N (n_0,\sigma_0) = \sum_{n=0}^\infty e^{-\frac{(n-n_0)^2}{2\sigma_0^2}}
\ee
is the normalisation factor. The parameter $\sigma_0$ represents the standard deviation of the Gaussian form of the coefficients and of the normalisation factor; $n_0$ is a large quantum number which is selected in such a way that the energy is large. Finally, $\phi_0$ is a parameter that is related to the initial conditions (and in numerical calculations it is constrained by the initial conditions $p_0=p_{n_0}$ and $x_0 = -\frac{L}{2}$).

Any quantum state that is expanded as in equation \eqref{expansion} satisfies the boundary conditions trivially. This is the case for the Gaussian Klauder coherent state with the above-mentioned coefficients. Therefore our quantum state is indeed an element of the Hilbert space. To ensure the macroscopic character of our state, we select the parameter values so that our lengths and energies are much larger than $E_0$. In the appendix \ref{quasi-classical}, we present a detailed analysis of the behaviour (dynamics) of the quantum state in the infinite square well in this quasi-classical and macroscopic limit so that we understand better the parameter space and to ensure that we are in the correct limit. Here we present in Figures \ref{fig:sigma10_psi} and \ref{fig:momentum_space} the form of the coherent state at the initial time in the position and momentum space. It is shown that the probability distribution of the wave function is initially localised both in space around $x_0=-\frac{L}{2}$ and in momentum around $p_0$ as expected.

Since we have justified that our system describes a quasi-classical and macroscopic system, we would not expect to observe any of the quantum paradoxes that arise when considering microscopic systems. We proceed by giving the two different coarse-grainings that, within the consistent histories formulation, lead to contrary inferences for this quasi-classical and macroscopic system. Specifically, we pose the question of the time-of-arrival problem for a quasi-classical particle represented as a suitably localised wave packet initially located at $x = -\frac{L}{2}$ with positive average momentum $p=p_0$ and confined in an infinite square well $[-L,L]$. We deal with the question on whether it passes $x=0$ or not by considering the complementary question of whether the particle always remains in the subregion $[-L,0]$ and we construct ``Zeno'' histories \cite{Wallden:2007aa}. 

\subsection{Construction of the coarse-grainings: The first consistent set}
\label{first}
\begin{figure}[t]
\centering
\begin{subfigure}[t]{0.45\textwidth}
\centering
\includegraphics[height=4cm,width=6.5cm]{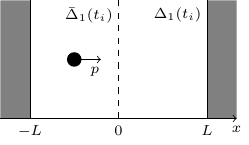}
\subcaption{The time-of-arrival problem in the infinite square well}
\label{fig:arrival_time1}
\end{subfigure}
\begin{subfigure}[t]{0.45\textwidth}
\centering
\includegraphics[height=4cm,width=6.5cm]{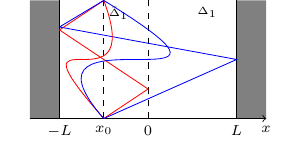}
\subcaption{Various histories}
\label{fig:histories1}
\end{subfigure}
\caption{Left Panel: The time-of-arrival problem in the infinite square well; this is an instant of the partition for the first set of histories. The two subregions are $\Delta_1=[0,L]$ and $\bar{\Delta}_1 = [-L,0]$. The question is whether the particle, equipped with positive momentum will be even for only one moment in $\Delta_1$. The barrier at $x=0$ is fictitious and corresponds to infinitely many projections with the operator \eqref{projection_continuous}. The projection takes place also at $x=-L$ but this also coincides with the real barrier of the square well.  \newline Right Panel: Various histories: the blue set belongs to $h_1$ while the red in $\bar{h}_1$}
\label{first_set}
\end{figure}
\paragraph{Coarse-graining and histories.}
For the first coarse-graining we define the intervals $\Delta_1 = [0,L]$ and $\bar{\Delta}_1 = [-L,0]$, as intuitively indicated by the question posed and group the possible paths into two sets to construct the coarse grained histories. The first history $h_1$ that ``never arrives'' contains the paths that remained at all times within $\bar{\Delta}_1$; while the second $\bar{h}_1$ contains all the other histories, i.e. those that at least at some time entered in the region $\Delta_1$ (see Figure \ref{fig:arrival_time1}). In other words:
\begin{subequations}
\begin{align}
&h_1 = \{ x (t) \  | \  x (t)  \in \bar{\Delta}_1 \forall \ t \in [0,\tau]\} \\
&\bar{h}_1 =  \{ x (t) \  | \  x (t)  \notin \bar{\Delta}_1 \text{ for some } t \in [0,\tau]\}.
\end{align}
\end{subequations}
It is important to note here, that our conclusion is valid even if the consistency condition is satisfied for a single specific choice of $\tau$. In the following we will treat $\tau$ as variable and see for which choices of $\tau$ the condition is satisfied. Interestingly, we will find that the consistency condition is satisfied for large ``intervals'' so it is ``robust'' in time. For example a good choice for our aims is $\tau=0.85 T_{cl}$, in units of the classical period of the well $[-L,L]$, when the centre of the wave packet has returned to its initial position but with opposite average momentum.

\paragraph{Consistency condition}

\begin{figure}[t]
\centering
\begin{subfigure}[t]{0.45\textwidth}
\centering
\includegraphics[height=4cm,width=6.5cm]{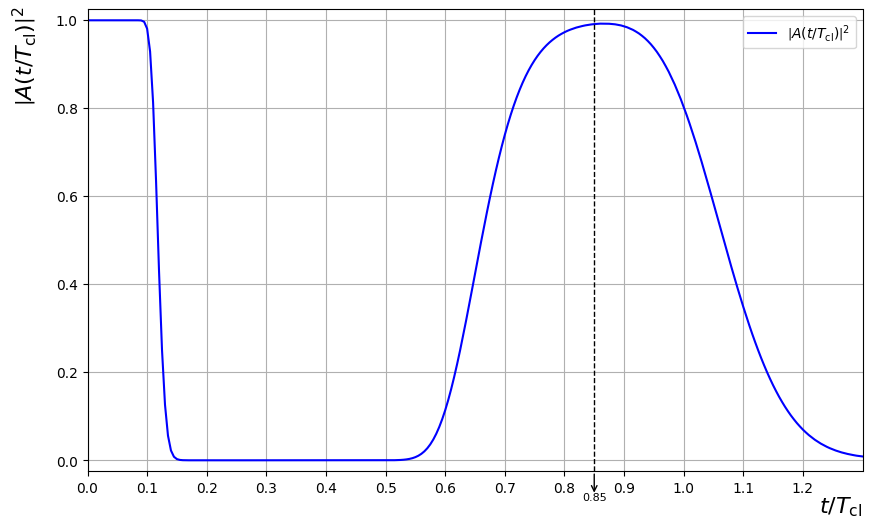}
\label{fig:overlap1_left}
\end{subfigure}
\begin{subfigure}[t]{0.45\textwidth}
\centering
\includegraphics[height=4cm,width=6.5cm]{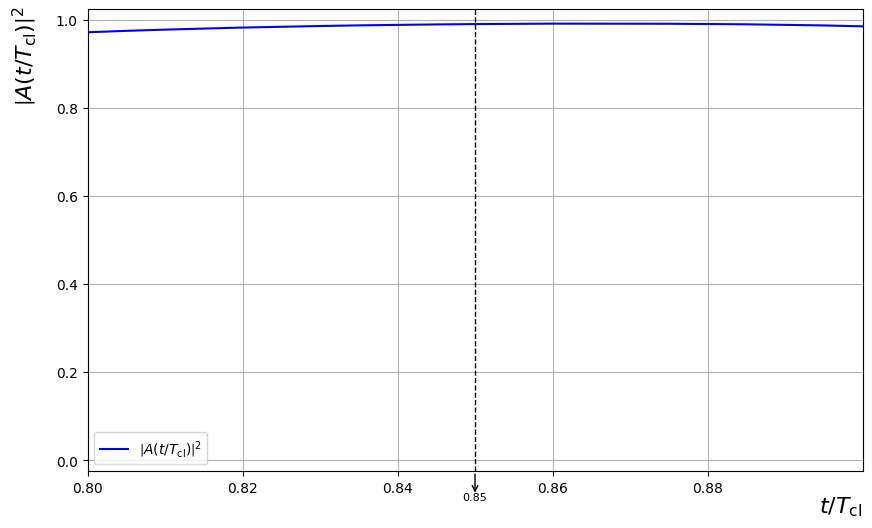}
\label{fig:overlap1_right}
\end{subfigure}
\caption{The square of the overlap of the two wave packets as a function of time normalised with the classical time (left plot) and the "zoomed" plot near $\tau = 0.85 T_{cl}$ (right plot). The parameters the same as follows $n_{0}^1 = 100$, $n_0 = 2 n_{0}^1 + 1$, $\sigma_0^1 = 10$, $\sigma_0 = 2 \sigma_0^1$. The phases $\varphi_0^1$ and $\varphi_0$ are selected such that both wave functions are localised at $x/L = -1/2$. The superscript $1$ refers to the coherent state in the well spanning $[-L,0]$.}

\label{fig:overlap1}
\end{figure}

To show that this set is consistent, we need to check the relation \eqref{consistentcond2} and the boundary condition \eqref{boundcond}. For the boundary condition to be satisfied we must have $\Psi (-L,0) = 0 $. This holds trivially since it is one of the boundaries of the $[-L,L]$, and we have caclculated that $\Psi (0,0) = 0$ up to the order $O(10^{-25})$. For the condition \eqref{consistentcond2} to hold we have to find at which times the overlap 
\be
\label{overlapA}
A(t) = \langle \bar{\Psi}_1 (x,t)\Psi (x,t) \rangle  = \int_{- L}^0 {dx} \bar \Psi_1(x,t)\Psi(x,t)
\ee
of the quantum state and its restriction in $\bar{\Delta}_1$ is one. As Figure \ref{fig:overlap1} shows, the overlap starts increasing at $t\approx 0.5 T_{cl}$, reaches the value one at $\tau \approx 0.85 T_{cl}$ and decreases until $t\approx 1.25 T_{cl}$ after which is zero. The coincidence of the wave packets is in the interval $[0.75,0.95]$ of the classical period is actually expected from the coincidence of the classical trajectories of the full and restricted evolution of the wave packets. Therefore, we can see that it is an extended time interval around the value $\tau = 0.85 T_{cl}$ that the overlap admits its maximum value of $1$. Thus the consistency condition is quite robust in time. Therefore, since both conditions hold, at least approximately, we have shown that this set of histories is consistent and we can assign probabilities to the histories.

\paragraph{Probabilities}
The candidate probability, for remaining during all the interval $[0,\tau]$ in the region $\bar{\Delta}_1$ is given by the expression $D(h_1,h_1)=\mu(h_1)$ and it is not difficult to see that it is one for $\tau = 0.85 T_{cl}$:
\begin{equation}
    \mu(h_1)=\bra{\Psi_0}g_r^\dagger(\tau,0)g_r(\tau,0)\ket{\Psi_0}\approx 1 
\end{equation}
We have not written a strict equality because of the approximations and assumptions we made for calculating the consistency condition in the previous section. The consistency condition signifies the vanishing of the interference terms between coarse-grained sets of histories. If this is satisfied approximately, this will also be the case for the classical probability rules. 

This result, that mathematically is the quantum Zeno effect for the continuous variables $x$ and $p$, is well known. However, in general, this ``Zeno'' history does not decohere with its negation, and therefore the consistent histories cannot assign (this or another) probability to such history. After all the ``real quantum Zeno'' effect arises due to the act of continuous measurements, while in our case there is no observer making such measurements. In contrast, our example with the given choices of walls and choosing $\tau\approx 0.85 T_{cl}$, does decohere as we saw. Therefore unlike other ``Zeno-type'' histories we get:
\begin{eqnarray}
    \mu(\bar{h}_1)&=&\bra{\Psi_0}(g^\dagger(\tau,0)-g_r^\dagger(\tau,0))\times(g(\tau,0)-g_r(\tau,0)\ket{\Psi_0}\nonumber\\
    &=&\bra{\Psi_0}g^\dagger(\tau,0)g(\tau,0)\ket{\Psi_0}-\bra{\Psi_0}g_r^\dagger(\tau,0)g(\tau,0)\ket{\Psi_0}-\nonumber\\
    & & -\bra{\Psi_0}g^\dagger(\tau,0)g_r(\tau,0)\ket{\Psi_0}+\bra{\Psi_0}g_r^\dagger(\tau,0)g_r(\tau,0)\ket{\Psi_0}\nonumber\\
    &=&2-2 \Re \bra{\Psi_0}g_r^\dagger(\tau,0)g(\tau,0)\ket{\Psi_0}\approx 0
\end{eqnarray}
where for the last equality we used the previous result from the consistency condition. We can see that indeed $\{h_1,\bar{h}_1\}$ form a consistent set with probabilities
\begin{subequations}
\bal
&p(h_1) = 1, \\
&p(\bar{h}_1) = 1- p(h_1) = 0.
\eal
\end{subequations}
These show that the answer to the initial question of the particle crossing the origin within the time interval $[0,\tau]$, is negative.
\subsection{Construction of the coarse-grainings: The second consistent set (time-dependent)}\label{second}
\begin{figure}
\centering
\begin{subfigure}[t]{0.45\textwidth}
\centering
\includegraphics[height=4cm,width=6.5cm]{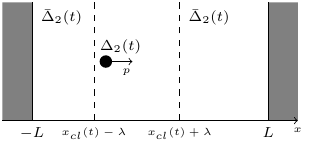}
\subcaption{Instant of the coarse-graining for the second histories set.}
\label{fig:arrival_time2}
\end{subfigure}
\begin{subfigure}[t]{0.45\textwidth}
\centering
\includegraphics[height=4.5cm,width=6.7cm]{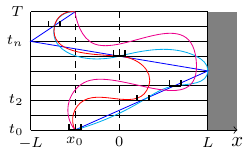}
\subcaption{Various histories}
\label{fig:histories2}
\end{subfigure}
\caption{Left Panel: Instant of the coarse-graining for the second histories set. In this case, the partition is $\Delta_2 (t)=[x_{cl} (t) -\lambda, x_{cl} (t) + \lambda]$  and changes in time, following the classical path of the particle. The two barriers move uniformly and continuously in the infinite square well due to the quasi-classical nature of the problem, see section \ref{discussion} and this one for detailed discussion on the continuity. \newline Right Panel: Various histories: the blue set belongs to $h_2$ while the red in $\bar{h}_2$}
\label{second_set}
\end{figure}
\paragraph{Coarse-graining and histories}
Here we are interested in histories that remain within a (time-dependent) interval that at all times contains the centre of the wave packet and is sufficiently wide to include essentially the full wave packet. One can define such a time-dependent interval in different ways and care needs to be taken when the classical trajectory (and the centre of the wave packet) hits one of the walls placed at $\pm L$. The first important point is that this interval should be much larger than the width of the wave packet at all times, and this can be achieved if the moving interval is $O(L)$ since the length of the well $L$ is significantly larger than the width of the Gaussian wavepacket in the position space which in our case it is true. The approach we will take is that at each moment of time we will apply the position projection operator 
\be
P_2 (t)= \int_{\Delta_2 (t)} dx \ketbra{x}{x}
\ee 
where the (time-dependent) interval $\Delta_2 (t)=[x_{cl} (t) -\lambda, x_{cl} (t) + \lambda]$ follows the trajectory of the wave packet in the well. We will take $\lambda$ to be the width of the interval and its centre is the instantaneous mean position of the wave packet or, equivalently, the classical trajectory, $x_{cl} (t)$. An instant of this situation is depicted in Figure \ref{fig:arrival_time2}. 

Following the same procedure as described before, the interval  $\Delta_2$  is constructed by very frequent application of the position projection operator $P_2 (t)$, as a ``Zeno'' box. At each moment of time, the position of $\Delta_2 (t)$ changes since it follows the motion of the particle. 

We take the interval $\Delta_2 (t)$ to have width $\lambda = L/8$ and the mathematical effect of these projections is to introduce ``fictitious'' infinite potential walls at the boundary of this region.
It is clear that when $x_{cl}(t)+L/8>L$ or $x_{cl}-L/8<-L$, the fictitious wall is further than the real walls that exist in $\pm L$. This means that the box (infinite square well) that the restricted evolution takes place, in those cases, changes size to reach $L/8$ when $x_{cl}(t)=L$ from the initial $L/4$ width. Since the length of the well is significantly larger than the width of the Gaussian wavepacket this does not change significantly the analysis. Alternatively, one could use a different interval that ``freezes'' when $x_{cl}(t_1)+L/8=L$ and then starts following the wave packet at $t_2$ when the centre has returned at $7L/8$ but is moving the opposite direction. We focus on the first case, but the consistency condition and probabilities would be the same in the second case too. 

The interval $[-L,L]$ is divided in three subregions; we are interested in the histories in the second one, those that never leave $\Delta_2$. These are constructed by all the paths reflected from the boundaries of the interval.  
In this way, the set of histories belonging in the interval $\Delta_2 (t)$ are: 
\begin{subequations}
\bal    
&h_2 =  \{ x (t) \  | \  x (t)  \in \Delta_2 (t) \ \forall t \in [0,\tau]\}
\eal
The contributions of the paths outside $\Delta_2$ constitute the set of histories that do not belong to $\Delta_2$ at least for some moment, that is
\bal
&\bar{h}_2 =\{ x (t) \  | \  x (t)  \notin \Delta_2 (t) \text{ for some } t \in [0,\tau]\}.
\eal
\end{subequations}
\paragraph{Consistency condition} 
Now, we have to establish that the consistency condition is satisfied, i.e. that the initial wave function vanishes at the boundaries of $\Delta_2$ and that the overlap between the restricted and full solution of the Schr\"odinger equation is one initially as well as at later times, since the interval $\Delta_2$ is time-dependent. 

It can be seen that at the initial moment $t_0=0$, the overlap between the wave packet defined in $\Delta_{2L}$ and its restriction in $[x_{cl}-\lambda,x_{cl}+\lambda]$ is equal to one:
\be
\int_{x_{cl}-\lambda}^{x_{cl}+\lambda} \Psi^* (x,t_0) \Psi_2 (x,t_0) dx \approx 1
\ee
because we have selected the interval $\Delta_2 (t_0)$. The translational symmetry guarantees that this value will be approximately preserved in time while it moves in the infinite square well because of the quasi-classical properties of the Gaussian state. Viewing this differently, the unrestricted evolution of our quasi-classical system does not spread and keeps the same width in position and momentum space, while the centre of the wave packet follows the classical trajectory. By construction the same holds for the restricted evolution. Moreover, the wave packet bounces at the  walls $\pm L$ in both cases, while the fictitious walls of the restricted evolution are always at point that the wave function is vanishingly small because $L/8\gg $ is much larger than the width of the wave packet.
\paragraph{Probabilities}
Through the above procedure we have thus constructed a consistent set, since it satisfies the two sub-conditions into which the consistency conditions breaks into: i) the boundary condition and ii) the overlap between the full and restricted evolution of the wave packet. We can now calculate  
the probability the particle is always in $\Delta_2 (t)$. By construction, the particle never leaves the subspace $\Delta_2 (t)$, thus the probability to find it there is one:
\begin{subequations}
\bal
&p(h_2) = 1, \\
&p(\bar{h}_2) = 1- p(h_2) = 0
\eal
\end{subequations}
What is crucial to note here is that all the paths that remained through all the interval $[0,\tau]$ in $\Delta_2(t)$ passed through $x=0$. This is true since, for example, at $t=T_{cl}/4$, the interval $\Delta_2(t)=[3L/8,5L/8]$ and is fully in the positive axis meaning that paths that at that moment are within $\Delta_2(t)$ have been in the $x>0$ region. In other words $h_2$ is a strict subset of the histories that crossed $x=0$. We can therefore infer from this consistent set, that the particle passed $x=0$ with certainty (probability unity). Interestingly, while this consistent set gives an indirect way to answer the time-of-arrival question, it actually recovers the answer that one would intuitively expect.

\section{Discussion}\label{discussion}
In this work, we considered the construction of two different consistent sets and showed that \emph{it is possible to have contrary inferences even at the quasi-classical limit within the consistent histories formalism}. In standard quantum theory, contextuality related paradoxes exist only in the microscopic/quantum regime and these do not appear when the system has decohered with respect to a quasi-classical basis. In particular, as explained in section \ref{time_arrival}, contrary inferences occurring with certainty are impossible in single-time quantum theory, but possible in the histories setting. However, we showed that for observable quantities (position histories) extended-in-time, the quantum paradoxes, and especially those related with contrary inferences, persist at the quasi-classical limit.
Therefore in the consistent histories, the context/consistent set dependence is more significant and points to the need of supplementing the formalism with certain selection criteria.

The two scenarios we considered were: \begin{enumerate}[(i)]
\item The system does not leave the subregion $\bar{\Delta}$.
\item The system remains always within a small box that follows the classical evolution of the centre of the classical object/wave packet. 
\end{enumerate}
We considered a particle in the infinite square well, initially positioned at the negative axis and moving to the right and asked whether the particle will pass from the origin $x=0$ at some moment $t \in [0,\tau]$. The particle was quasi-classical and macroscopic (a coherent state) with expectation value of the position at $x=L/2$ and positive expected value for the momentum. We selected two different coarse-grainings corresponding to the situations discussed above. More specifically, we constructed the following histories:
\begin{align*}
&h_1 = \{ x (t) \  | \  x (t)  \in \bar{\Delta}_1 \ \forall t \in [0,\tau]\} \\
&\bar{h}_1 =  \{ x (t) \  | \  x (t)  \notin \bar{\Delta}_1 \ \text{ for some } t \in [0,\tau]\}\\
&h_2 =  \{ x (t) \  | \  x (t)  \in \Delta_2 (t) \ \forall t \in [0,\tau]\}  \\
&\bar{h}_2 =\{ x (t) \  | \  x (t)  \notin \Delta_2 (t) \ \text{ for some } t \in [0,\tau]\}
\end{align*}
that belong in the histories space $\Omega$ and pairwise ($\{h_1,\bar{h}_1\}\ ; \ \{h_2,\bar{h}_2 \}$) form different partitions. In sections \ref{first} and \ref{second}, we demonstrated that the consistency conditions \eqref{boundcond} and \eqref{consistentcond2} are satisfied at least under approximations necessary to take the quasi-classical limit. 
We thus found the following probabilities for the histories:
\begin{align*}
&p(h_1)= D(h_1, h_1) = \mu(h_1) =1,\label{measure_h1} \\ 
&p(\bar{h}_1)=D(\bar{h}_1,\bar{h}_1) = \mu (\bar{h}_1) =0, \\ 
&p(h_2)=D(h_2,h_2 )= \mu (h_2)=1, \\
&p(\bar{h}_2)=D(\bar{h}_2, \bar{h}_2) = \mu (\bar{h}_2) = 0 
\end{align*}
We emphasise that these were the {\it candidate} probabilities, which by virtue of the consistency of the sets ($D(h_i,\bar{h}_i) =0, \ i=1,2$), came to satisfy the rules of the classical probability theory; that is $\mu(h_i) + \mu(\bar{h}_i) =1, \ i=1,2$. We also mention that these probabilities are valid for the time interval close to $\tau = 0.85 T_{cl}$.

We now show how contrary inferences appear in this setting. 
First, we recall that we have contrary inferences when two (coarse-grained) histories have both zero quantum measure and cover the histories space $\Omega$.  
In our example zero cover of $\Omega$ is the set $\{ \bar{h}_1, \bar{h}_2\}$. Their quantum measure is zero, while it is clear that any trajectory belongs to either $\bar{h}_1$ or $\bar{h}_2$ or to both sets, and thus it forms a cover. Second, we also have that $\mu (h_1) = 1$ and $\mu (\bar{h}_2) =0$. If we assign a truth value $\mathsf{True}$  to events/histories with probability one and $\mathsf{False}$ to events/histories with probability zero, we are led to contrary inferences because of the modus ponens (implication inference) which states that if $A=\mathsf{True}$ and $A\implies B$ (i.e. subset) jointly implies $B=\mathsf{True}$ something that does not hold in our example.

It is important at this point to emphasise that the initial motivation of the consistent histories formalism is the description of closed quantum systems without measurements or external observers. This is achieved by replacing the process of measurement with the consistency condition that must be satisfied in order for a question to be (classically) answered. As it was demonstrated, the same question can be answered by considering different partitions of the histories space. For each of those partitions, to calculate the amplitudes and their interference one needs to use projections and/or restrictions in the path integral. Mathematically the different partitions appear as if different measurements took place, but in consistent histories there is no ``real'' measurement taking place, or external observer.  The physical system (paths and true Hamiltonian) are identical in all partitions, therefore reaching a different conclusion about properties of the same classical system seems problematic and gives further evidence that consistent histories without a set-selection mechanism may not be a viable alternative.
\acknowledgements{}
P.W. thanks Fay Dowker and Rafael Sorkin for early discussions and acknowledges support by the UK Hub in Quantum Computing and Simulation with funding from the EPSRC grant EP/T001062/1. A.Z. acknowledges support by the UNAM Postdoctoral Program (POSDOC) and by the National Science Centre in Poland under the research grant Maestro (2021/42/A/
ST2/00356).



\appendix \label{appd}

\section{Preliminaries}
\label{Preliminaries}
\subsection{The consistent histories formalism}
\begin{figure}[t]
\centering
\includegraphics[height=5.5cm,width=8cm]{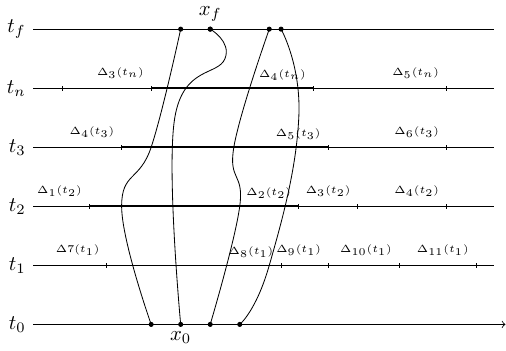}
\caption{Paths of a non-relativistic particle on the two-dimensional spacetime diagram. The particle starts from a point $x_0$ at $t_0$ and ending at some $t_f$ after some time $t_f$. In every intermediate time $t_1, t_2,...,t_n$ there is a spatial partition (exhaustive and exclusive) depicted by the intervals $\Delta_i (t_j)$.}
\label{fig:paths_spacetime}
\end{figure}
In this section, we briefly review the histories formalism for completeness and to fix the notation. We are dealing with a closed quantum system, with no external environment\footnote{In cases that we have a quantum system interacting with some, apparently, external environment, we simply include the degrees of freedom of the ``environment'' in the definition of our quantum system.} or observer. The physics is fully captured by the kinematics and dynamics of the quantum system. The kinematics defines the space of possible histories, similarly to the space of possible ``paths'' in the path integral view of quantum theory. The dynamics is determined by the Hamiltonian (or the action) of the system, and mathematically enter the picture with the decoherence functional (see later) that determines the ``amplitudes'' of different histories and their ``interference''. It is well known that due to interference, one cannot in general assign classical probabilities to those histories (c.f. double slit). The central aim of the consistent histories is to find under what conditions can someone assign classical probabilities to (coarse-grained) sets of histories. The main virtue of the approach is that those conditions are derived from the quantum system (dynamics and kinematics) without any further assumption about observers and/or measurements.

A history, that we will denote as $h_i$,  
is a generalised object that gives a description of the system at different times. A finest-grained history gives a complete description of the system at all moments of time. 
The set of all finest-grained histories comprises the histories space $\Omega$. For instance, in the case of the non-relativistic particle, fine-grained histories are all the single-valued functions $x^{i} (t)$ on the configuration space of the coordinate variables $(t, x^i)$ that the particle can follow under certain boundary conditions. We will use this example to clarify concepts in our brief introduction since it is the most relevant for our work. In general a (coarse-grained) history $h_i$ corresponds to a set of fine-grained histories (in our example a set of trajectories $x^i (t)$).
The space of histories $\Omega$ can be partitioned into a collection of coarse-grained histories $\{h_i\}$ that are exhaustive $\cup_{i} h_i= \Omega$ and disjoint $h_i \cap h_j=\varnothing \ \forall \ i,j$. We call such set of histories a \emph{partition}. There is no unique way to partition the space of histories, and as we will see not all partitions are suitable for predictions. Histories, as quantum objects, interfere and in general it is not possible to assign consistently a classical probability to a particular (fine-grained or coarse-grained) history. Instead, one can define a bi-linear function on the space of histories, called the decoherence functional
\be
D (h_i,h_j) = \int_{x^i(t)\in h_i} \mathcal{D} x^i \int_{x^j(t)\in h_j} \mathcal{D} x^{j} \exp \{ i \left(S[x^i] - S[x^j\right) \}\delta(x_f^i - x_f^j) \rho (x_0^i, x_0^j)
\ee
Here again we use the example of a non-relativistic particle for simplicity. The decoherence functional is a complex-valued, bi-linear function defined on the space of all (fine-grained or coarse-grained) histories $h_i, h_j$. The integral is defined over the paths $x^i(t)$ in the coarse-grained histories $h_i$ and $h_j$ respectively. The paths start from $x_0$ at $t_0$ and end at $x_f$, and are defined within the time interval $[t_0,t_f]$, while $\rho (x_0^i, x_0^{j})$ is the initial density matrix; the $\delta$-function ensures that histories with different end-point decohere (we can mathematically think of having a final-time position measurement).

The decoherence functional essentially measures the interference between the histories. When applied to a partition of the histories space, it measures interference of different answers within the context defined by the partition/coarse-graining. The decoherence functional satisfies the following properties:

\begin{enumerate}[(i)]
\item hermiticity: $D(h_i,h_j) = {D^\dag }(h_j, h_i)$
\item positivity: $D(h_i, h_j) \geq 0$
\item normalisation: $\sum_{h_i, h_j}D(h_i, h_j)=1$
\item superposition principle $D(h, h')=\sum_{\substack{h_i \in h, \\ h_j \in h'}} D(h_i, h_j)$
\end{enumerate}
An important special case is when one considers a partition (recall  $\cup_i h_i=\Omega\ ; \ h_i\cap h_j=\emptyset$), such that the non-diagonal elements of the decoherence matrix vanish, that is 
\be\label{generic_cons_cond}
D(h_i, h_j) =0, \ \forall \ i \neq j
\ee
Then the (coarse-grained) histories that make this partition, form a \emph{consistent set} of histories, and
the diagonal elements of the decoherence functional give the probabilities of the corresponding histories to happen, that is
\be\label{candidate_prob}
p(h_i) := D(h_i, h_i).
\ee
The relation \eqref{generic_cons_cond} 
is known as the consistency condition and it is the key mathematical structure in consistent histories, giving us a tool to identify when a coarse-grained set of histories (partition) of a closed quantum system admits consistent classical probabilities.

The expression of \eqref{candidate_prob}, when applied to a general history (not one belonging to a consistent set) is usually referred to as \emph{candidate probability}. If a history belongs to a consistent set (i.e. at a partition obeying \eqref{generic_cons_cond} ensuring that alternatives within that set have no interference), the candidate probability is the actual probability of that history, while in general, it cannot be treated as probability because of interference.
The decoherence functional and the related candidate probabilities inspired a generalisation of standard measure theory to quantum measure theory \cite{sorkin_quantum_1994}, a mathematical construct that replaces Kolmogorov sum rule with a weaker higher-order rule $\mu(A\cup B\cup C)=\mu(A\cup B)+\mu(A\cup C)+\mu(B\cup C)-\mu(A)-\mu(B)-\mu(C)$. It can be shown that the diagonal elements of a decoherence functional always define a quantum measure (or more precisely a ``level-2 measure'', since higher level measures were also defined):
\be\label{quantum_meas}
\mu (h_i): = D(h_i, h_i).
\ee
In the following we will be using quantum measure notation to denote the diagonal elements of the decoherence functional.

In the non-relativistic quantum mechanics, the decoherence functional can also be formulated as a sequence of time-ordered projection operators that represent physical events at different moments of time
\be
D(h,h') = \Tr (C_h^\dagger \rho C_{h'})
\ee 
where 
\be
C_h = P_{h_n}U(t_n-t_{n-1})...U(t_2 - t_1) P_{h_1} (t_1)
\ee
$U(t) = \exp(-i Ht)$ is the unitary evolution operator, $H$ the Hamiltonian of the system and $P_{h_i}$ is the projection operator on the subspace that the history $h_i$ lies at time $t_i$.

Note that the operator and path-integral descriptions are not always equivalent since they start from different points. For non-relativistic quantum mechanics the Hamiltonian formulation is more general (e.g. the projections at a given time can be either configuration space or momentum space) but both approaches are known to be equivalent for sets of histories defined by paths on the configuration space \cite{Hartle1993} as in our example.
In this work, projections at any given time are on different ranges $\Delta(t)$ of the position that in general depend on the time instance
\be\label{projection_continuous}
P = \int_{\Delta(t)} dx \ketbra{x}{x}
\ee
An important issue in the consistent histories approach is that there exist many different partitions (sets of histories) that decohere (satisfy the consistency condition) and do not arise as different coarse-grainings of a single finest-grained consistent set. This could become problematic when trying to infer what happened comparing predictions from incompatible consistent sets. Here it is important to note that all consistent sets correspond to the very same quantum system -- the physics (e.g. Hamiltonian) is identical while there is no measurement or observer involved to differentiate the scenarios. It is simply a different view of the same system, each with equal importance, unless the standard formalism of the consistent histories is supplemented with some extra structure (e.g. a criterion selecting a preferred consistent set).

A final point to note is that in real systems coarse-grained histories rarely decohere exactly. Instead one deals with \emph{approximate} decoherence i.e. partitions that obey the consistency condition up to some (negligible) $\epsilon$. This is not considered a major issue for two reasons. First it is difficult to test observables in arbitrary precision and it seems unlikely that some negligible interference between histories would alter any observable quantity. Second it has been shown that for any approximately consistent set, it is possible to find an \emph{exact} set that is close in a suitably defined sense \cite{dowker1995properties, dowker1996consistent}. 


\section{Quasi-classical and macroscopic approximation\label{quasi-classical}}
Here we present an analysis of the behaviour (dynamics) of the quantum state in the infinite square well in the quasi-classical and macroscopic limit so that we understand better the chosen parameters and to ensure that we are in the correct limit. 

The quantum state \eqref{expansion} satisfies trivially the boundary conditions of the infinite square well since any linear combination of the eigenstates will satisfy them. To take the quasi-classical and macroscopic limit, we have to select the  parameters such that $1 \ll {\sigma _0} \ll {n_0}$, see ref. \cite{fiset2014equivalent} and perform the summations within a range $[n_0-n_f, n_0+n_f]$ where $n_f$ is properly chosen so that the summation converges. In fact, because the energy quantum number $n_0$ is large, the summations can also be turned to integrals in the corresponding interval. Moreover, the wave function in the momentum space defined as:
\be
\Phi (p,t) = \int\limits_{ - L}^L {\frac{{dx}}{{\sqrt {2\pi \hbar } }}} {\Psi }(x,t){e^{ - \frac{{ipx}}{\hbar }}}
\ee
can be calculated numerically and we have proven that it has a Gaussian form (see Fig. \ref{fig:momentum_space}).

The parameter $\phi_0$ is a phase related to the choice of the initial conditions. We have thus chosen it properly so that the wave packet is localised at $x_0=-L/2$ and $p=p_0$ at $t=0$.

\begin{figure}[t]
\centering
\begin{subfigure}[c]{0.45\textwidth}
\centering
\includegraphics[height=4cm,width=6.5cm]{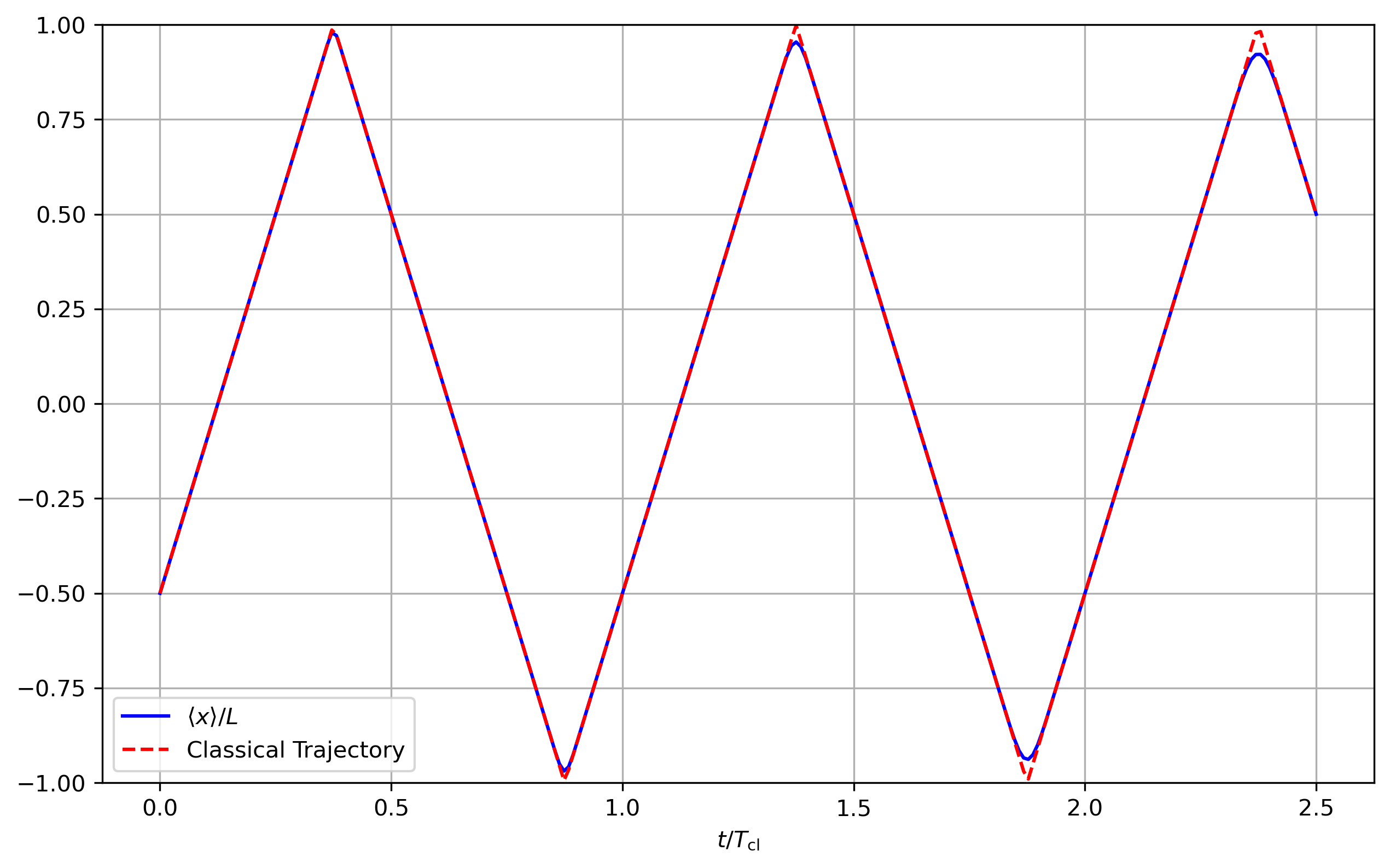}
\subcaption{Quantum vs classical trajectory.}
\label{fig:classical_trajectory}
\end{subfigure}
\begin{subfigure}[c]{0.45\textwidth}
\centering
\includegraphics[height=4.5cm,width=6.7cm]{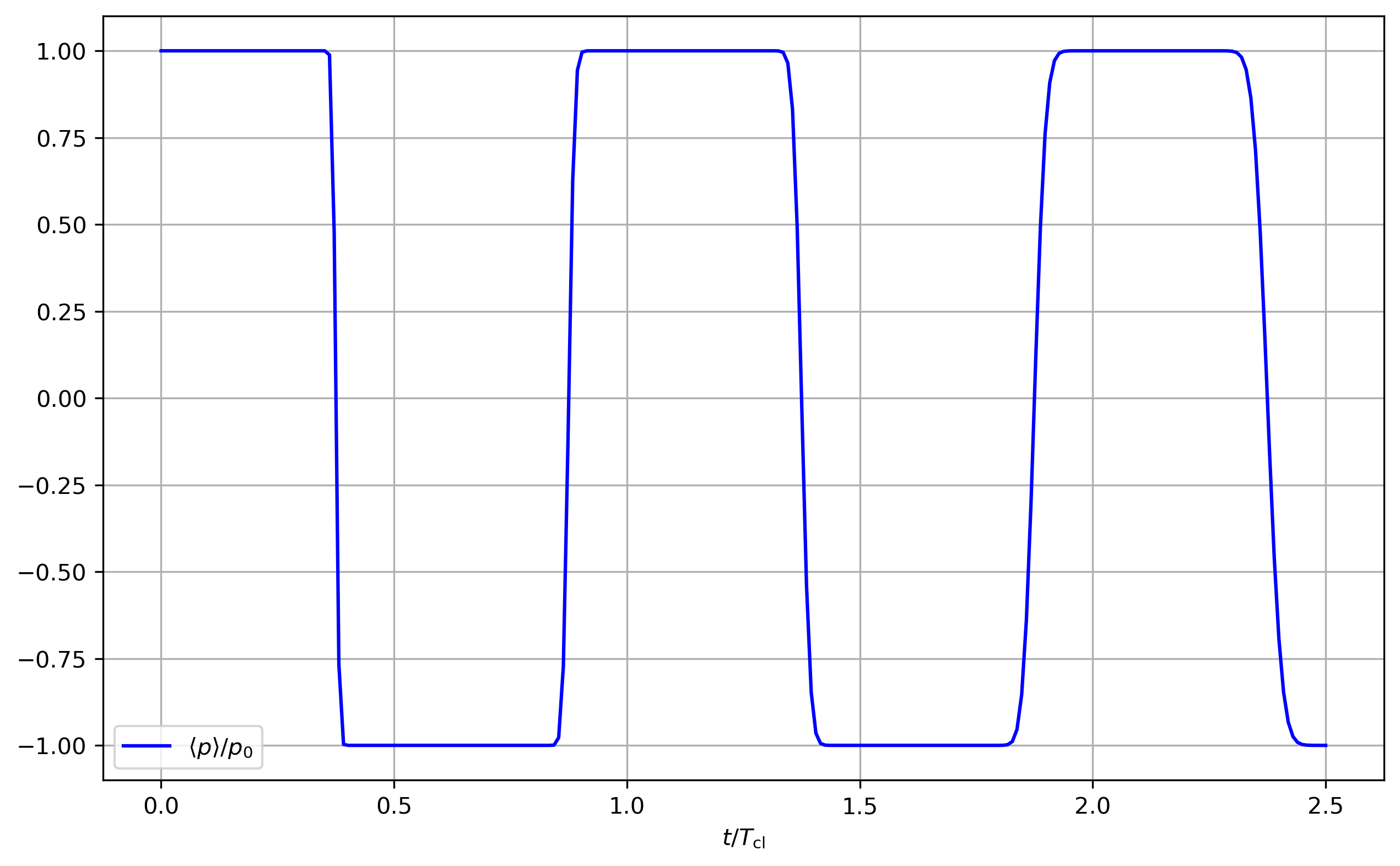}
\subcaption{Quantum Momentum.}
\label{fig:momentum}
\end{subfigure}
\centering
\begin{subfigure}[c]{0.45\textwidth}
\centering
\includegraphics[height=4.5cm,width=6.7cm]{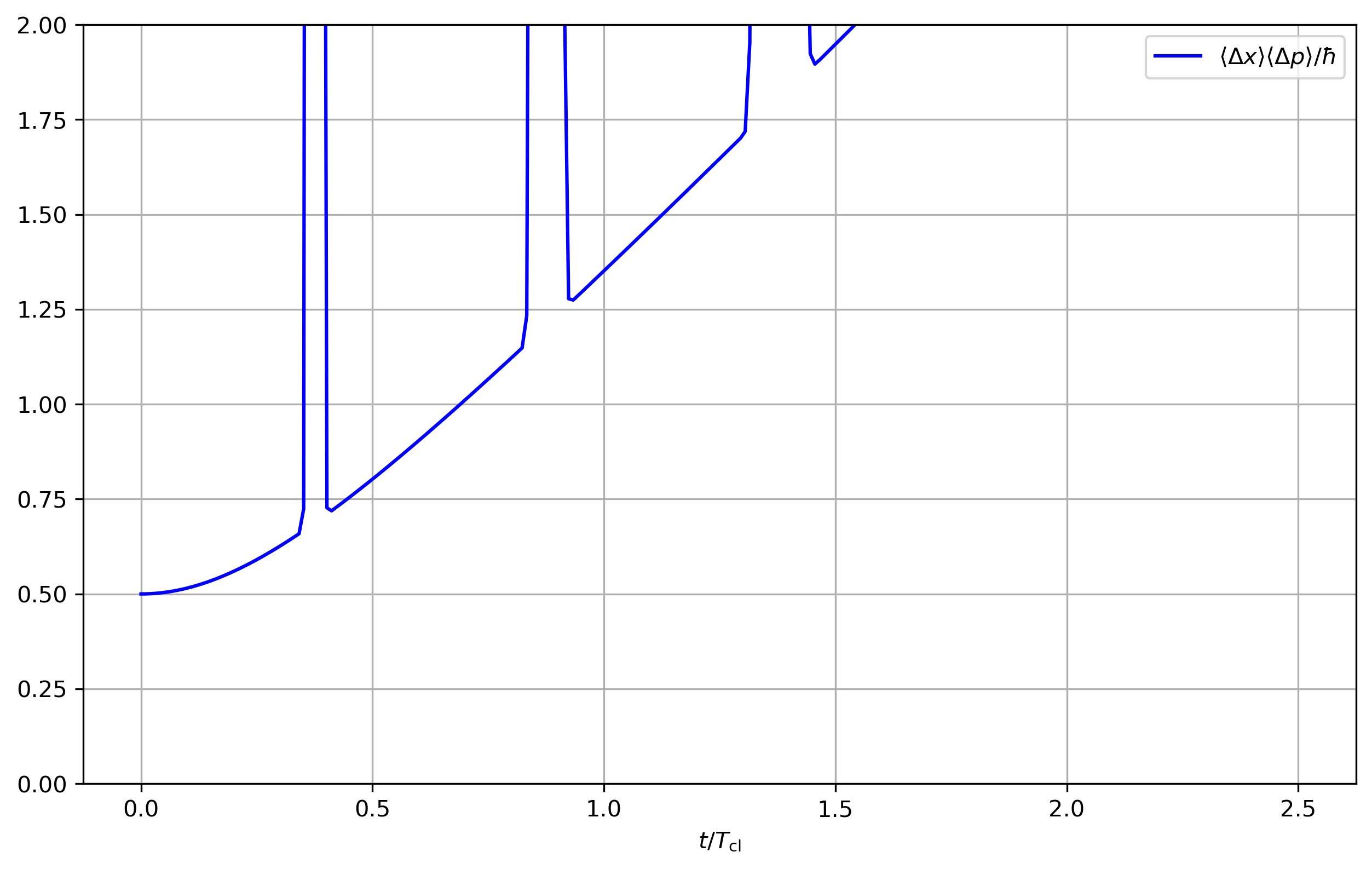}
\caption{Uncertainty Relation.}
\label{fig:uncertainty relation}
\end{subfigure}
\caption{Classical vs quantum behaviour of the wave packet.}
\end{figure}
Using the same parameters as the ones used in figure \ref{initial_prob_distr}, we have numerically verified the following:
\begin{enumerate}
\item The quantum mean value of the position defined as
\be
{\langle x \rangle } = \int\limits_{ - L}^L {dx} {{\bar \Psi }}(x,t)x\Psi(x,t)
\ee
follows the classsical trajectory for a few classical periods close to the initial time $t_0$, see figure \ref{fig:classical_trajectory}. 

The classical trajectory of a free particle in 1-dimension, with momentum $p_0$, starting position $x_0=-L/2$ moving between two walls at $\pm L$, is given by 
\begin{widetext}
\bal\label{classical_paths}
x_{cl} (t)= 
\begin{cases}
&x_{\rightarrow}^{(n)} \equiv -2 (2n) L + (x_0 +\frac{p_0}{M}t), \quad t_{2n} \leq t \leq t_{2n +1}, \ n=0,1,2...\\
&x_{\leftarrow}^{(n)} \equiv  2 (2n -1) L - (x_0 + \frac{p_0}{M}t), \quad t_{2n-1} \leq t \leq t_{2n} \ n=1,2,...
\end{cases}
\eal
\end{widetext}
where $t_n= n T_{cl}$, with $T_{cl}$ as defined above, being the classical period of the system while $x_{\leftrightarrows}$ denotes the direction of the motion of the particle. The above equation is used in the calculation of the classical trajectory as depicted in figure \ref{fig:classical_trajectory}.

\item The quantum mean value of the momentum defined as follows
\be
{\langle p \rangle } =  - i\hbar \int\limits_{ - L}^L {dx} {\bar \Psi}(x,t)\frac{{\partial \Psi(x,t)}}{{\partial x}}
\ee
also follows the classical one, see figure \ref{fig:momentum}. The expectation value of the momentum retains the initial value and changes direction as expected when the particle hits the walls.

\item The uncertainty relation is defined as:
\be
{\langle {\Delta x} \rangle }{\langle {\Delta p} \rangle } = \sqrt {{{\langle {{x^2}} \rangle }} - \langle x \rangle ^2} \sqrt {{{\langle {{p^2}} \rangle }} - \langle p \rangle ^2} 
\ee
where
\be
{\langle {{x^2}}\rangle } = \int\limits_{ - L}^L {dx} {\bar \Psi}(x,t){x^2}\Psi(x,t)
\ee
and
\be
{\langle {{p^2}} \rangle } = {\hbar ^2}\int\limits_{ - L}^L {dx} {\bar \Psi}(x,t)\frac{{{\partial ^2}\Psi(x,t)}}{{\partial {x^2}}}
\ee
and is plotted in Figure \ref{fig:uncertainty relation}. The uncertainty relation for some classical periods is close to $\hbar/2$ but after a few periods this picture is lost. The fact that the uncertainty principle does not satisfy the minimal relation (contrary to what happens for the harmonic oscillator) is well known in the case of the coherent states of the infinite square well e.g. \cite{fiset2014equivalent}.

\end{enumerate}
We end this appendix by commenting how we re-scale the variables to make them dimensionless for the numerical calculations presented in this paper. Every position variable ($x$, $\sigma_0$, etc.) is defined as a ratio of $L$. The momentum is defined as a ratio of the initial momentum. Finally, the time is defined as a ratio of the classical period $T_{cl}$ corresponding to the well $[-L,L]$.


\bibliographystyle{unsrtnat}
\bibliography{decoherent}

\end{document}